\documentclass[sigconf,nonacm]{acmart}
\AtBeginDocument{%
  }

\usepackage{longtable}

\begin{document}

\title{Co-designing a Preliminary Repository of Augmented Reality Concepts for Real-Time Emotion Regulation}

\renewcommand{\shorttitle}{Co-designing a Repository of AR Concepts for Real-Time Emotion Regulation}

\author{Graciela Camacho-Fidalgo}
\orcid{0009-0009-2978-9408}
\affiliation{%
  \institution{Texas A\&M University}
  \city{College Station}
  \state{Texas}
  \country{USA}
}

\author{Edgar Rojas-Mu\~noz}
\orcid{0000-0001-6909-375X}
\affiliation{%
  \institution{Texas A\&M University}
  \city{College Station}
  \state{Texas}
  \country{USA}
}

\begin{abstract}

Augmented Reality (AR) can be a positive therapeutic approach to support mental health and emotion regulation. Although AR techniques for therapeutic support exist, there is no user-centered, expert-informed understanding of how real-time AR designs can support people in emotional distress without disengaging them from their ongoing activities. This lack of reusable design resources hinders the adoption of AR for mental health support. This paper addresses this gap by introducing a co-designed collection of AR interventions describing how this technique can support real-time emotion regulation. The repository was created following a two-phase participatory design process. Phase 1 recruited 40 anxiety-prone individuals and used the Nominal Group Technique to list ideas on how AR affordances could support emotion regulation. Phase 2 recruited 10 mental health professionals to organize these ideas into thematic clusters and assess their clinical feasibility. The resulting AR design repository, grounded in user perspective and clinical expertise, identifies eight thematic clusters and 106 design ideas. This work represents a first step towards the development of seamless real-time AR interventions for mental health.

\end{abstract}


\begin{teaserfigure}
  \includegraphics[width=\textwidth]{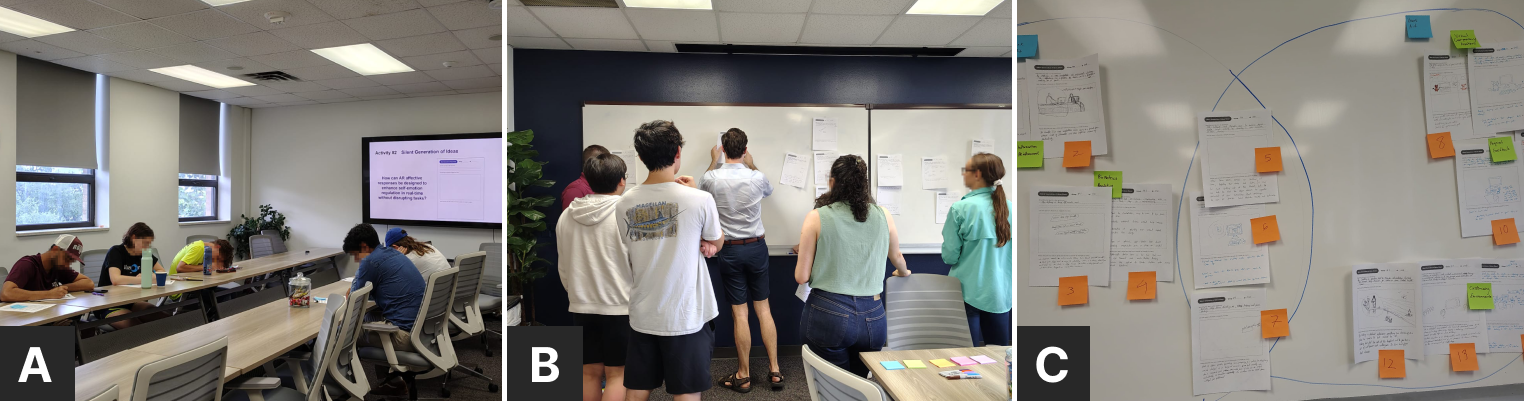}
  \caption{A two-phase participatory process was conducted to co-design a repository describing how AR can support real-time emotion regulation. Participants identified eight thematic clusters, each highlighting different affordances of AR technology for mental health, which were then evaluated by mental health professionals. Panels A-H depict ideas from these clusters that were ranked highly during the co-design process.}
  \label{fig:teaser}
\end{teaserfigure}

\maketitle

\section{Introduction}

In recent years, Augmented Reality (AR) has gained traction as a platform to support technology-mediated mental health and well-being interventions \cite{Sweileh2023}. Through AR's spatially-aware immersive graphics, patients can receive mental support in more engaging and interactive ways \cite{mittmann2022lina,bakir2023use}. Unlike other immersive techniques like Virtual Reality (VR), AR allows users to remain focused on their real-world surroundings, a grounding that can benefit mental health applications, e.g., less context switches, more situation awareness \cite{bakir2023use,rajkumar2024augmented}. Examples include practicing well-being techniques more consistently \cite{soler2024arcadia} and simulate stressful scenarios in controlled environments \cite{dalal2024revolutionizing}. Patients have reported multiple benefits from using AR, such as increased intrinsic motivation for well-being and longer commitments to their treatments \cite{rajkumar2024augmented}.

Despite AR's potential, its adoption in certain mental health contexts has outpaced its understanding. For instance, there is no user-centered, expert-informed understanding of how AR features can provide real-time, in-situ support for people in emotional distress without disengaging them from their ongoing activities \cite{lundin2023adverse,vial2022human}. Without these foundations, AR mental health interventions can struggle to foster emotional connection \cite{bakir2023use}, introduce unnecessary complexity \cite{cross2024digital}, and disengage users from their ongoing tasks \cite{borghouts2021barriers}. Ultimately, this lack of reusable design resources limits the adoption of AR for mental health support \cite{KIM2010669}.

This work addresses this gap by developing a repository of user-centered and expert-informed AR design concepts for real-time emotion regulation. The repository, built through a two-phase participatory co-design process, includes a collection of visual, auditory, and environmental ideas designed to support users' emotional states in real time, as shown in Figure~\ref{fig:teaser}. In the first phase, 40 anxiety-prone individuals participated in Nominal Group Technique (NGT) sessions to brainstorm ideas for AR-supported emotion regulation. In the second phase, ten mental health professionals organized these ideas into thematic clusters through a card-sorting activity and assessed their feasibility based on clinical criteria. The resulting repository represents an early conceptual resource to support future designs and developments of AR mental health applications.

The primary contribution of this work is a collection of AR design concepts for seamless real-time emotion regulation. We support this contribution through the following Research Objectives (ROs):
\begin{itemize}
\item \textbf{RO1}: Create a comprehensive list of user-proposed AR design concepts for real-time emotion regulation.
\item \textbf{RO2}: Organize the concepts  into thematic clusters based on their inherent characteristics.
\item \textbf{RO3}: Assess the perceived feasibility  of the clusters via mental-health professionals' feedback.
\end{itemize}

\section{Related Work}

\subsection{Technology-Assisted Mental Health Support}

 Mental health disorders affect more than 970 million people worldwide, with anxiety and depression being the most common conditions \cite{who2022mental}. The COVID-19 pandemic intensified these statistics, leading to a 25\% increase in global prevalence of anxiety and a 28\% increase in major depressive disorders \cite{who2022mental}. These disorders have also been associated with higher healthcare costs \cite{konnopka2020economic, konig2020excess} and disruptions in interpersonal and workplace relationships \cite{Burton2008}.

Traditional therapeutic approaches to address these disorders often combine an assessment of user needs \cite{hilsenroth2007alliance, hilsenroth2004alliance} with evidence-based techniques such as cognitive-behavioral strategies \cite{wenzel2016cbt}, mindfulness exercises \cite{BLANCK201825}, or exposure-based therapies \cite{abramowitz2019exposure}. These techniques can be practiced during therapist-guided sessions or independently through prescribed activities \cite{rosenthal2011favorite,keech2021journaling}.

However, barriers such as limited accessibility, cost, and difficulties maintaining these practices outside clinical settings remain significant challenges \cite{reardon2017barriers,kauppi2015perceptions}. These shortcomings have prompted a growing interest in digital mental health interventions \cite{Banos2022}, which can expand access \cite{Chandrashekar2018}, enhance engagement \cite{Tremain2020}, and complement existing therapeutic approaches \cite{Denecke2022}. A wide range of digital interventions for mental health have been explored, including mobile applications for cognitive behavioral therapy \cite{newton2020mobile}, conversational agents for emotional support \cite{tudor2020conversational}, and wearable sensing technologies for emotion recognition \cite{saganowski2022emotion}. These techniques have received positive appraisals from patients and experts, highlighting their potential to assist traditional therapeutic approaches \cite{philippe2022digital,smith2023digital}.

\subsection{AR for Mental Health Support}

AR technologies have emerged as a promising technology-mediated approach for mental health interventions \cite{usmani2022future}. By superimposing virtual imagery over the patient's view of the real world \cite{10.1145/3637355}, these techniques allow more immersive and embodied experiences. Early work utilized AR graphics for exposure therapy, allowing patients to confront fears in a controlled manner by overlaying virtual stimuli into their real environments \cite{zimmer2021effectiveness}. AR has also been used with patients with post-traumatic stress disorder (PTSD), enabling therapist-guided simulations to help process trauma-related cues while maintaining a sense of control \cite{10.1145/3706598.3713115}. Beyond clinical interventions, AR has been used to promote stress reduction \cite{jiang2022digital}, support mindfulness \cite{viczko2021effects}, guide breathing exercises \cite{blum2020vrbiofeedback}; display calming nature-inspired overlays \cite{viczko2021effects}, and support emotion regulation through gamified activities \cite{soler2024arcadia} and virtual playful companions \cite{norouzi2019walking}. These applications have been shown to significantly reduce anxiety and phobic responses \cite{botella2010treating,liu2022mindful}. Studies also reported that AR mediated-interventions increase patients' engagement with treatment \cite{baus2014moving} and the likelihood of sustained, daily use \cite{conroy2025integrated}.

Unfortunately, several challenges still surround the use of AR techniques in mental-health contexts, limiting their widespread adoption. For instance, these techniques often require users to pause their activities to engage with the technology \cite{laurie2016making}. This can be disruptive to individuals who struggle to schedule mental-health practices into their routines \cite{firth2017challenges} or in situations where stopping ongoing activities is not feasible \cite{yehuda2024improving}. Another challenge is the lack of real-time responsiveness, as people are unlikely to wait for an application to guide their breathing or pause a heated conversation to receive support \cite{fang2025social}. 
Design limitations have also been reported, such as failures to create strong emotional connections due to virtual stimuli feeling artificial or insufficiently personalized \cite{tabassum2025exploring}, steep learning curves due to non-intuitive interfaces \cite{kiourexidou2024exploring}, or automated guidance mechanisms assisting unreliably or at the wrong moment \cite{fang2025social}. Furthermore, some immersive systems are deployed as standalone applications without mechanisms for structured follow-up or clinician oversight, limiting their long-term therapeutic capacity \cite{cushnan2024clinicians}. Ethical considerations can also arise, as AR techniques should provide assistance without attempting to replace the therapist or create a technological dependency, and should complement users' daily routines rather than push them to deviate from them \cite{carlson2023virtual}.

\subsection{AR for Mental Health Design Resources}

Our thesis is that the lack of user-centered, reusable, and expert-informed design resources hinders the adoption of AR in the field of mental health. This work addresses this gap by creating a preliminary repository of non-disruptive real-time AR interventions for emotion regulation. To our knowledge, no work has tackled the creation of AR design concepts for mental health as its main contribution, let alone for emotion regulation. Developing such recommendations can significantly enhance AR's reproducibility and usability in mental health \cite{MendesSantos2022}, ensuring applications remain clinically-relevant \cite{peters2023wellbeing}. This repository could also provide a common reference to support future prototypes, as the current diversity of approaches makes them difficult to be compared \cite{herpertz2025developing}.

While no AR design framework for mental health exists, prior work in adjacent areas (e.g., HCI, telehealth, chatbots) can be used as a basis to create a repository of AR design recommendations for emotion regulation \cite{doherty2010design,langarizadeh2017telemental,lee2020hear}. Research confirms that AR is favored over VR for these purposes, as it allows users to integrate the virtual elements in their real surroundings and ease the transfer of learned skills to the real-world \cite{riva2016transforming,bakir2023use}. Experts also agree that user agency is paramount, as manual control reflects real-life decision-making and builds a sense of self-efficacy and self-reliance \cite{seals2021effects,pons2022extended}. Conversely, fully automated guidance risks being perceived as interruptive or intrusive \cite{fang2025social}. AR techniques should also feel safe \cite{lundin2023adverse}, relevant \cite{goulet2025ethical}, and useful \cite{rajkumar2024augmented}. Moreover, AR designs must be adaptable, allowing users to adjust appearances, environmental details, and difficulties \cite{pons2022extended,jurcik2024efficacy}. Multisensory methods should also be included to enhance the sense of presence, ranging from encouraging physical movement and gestures \cite{marto2024scope} to incorporating ambient sound and visual fidelity \cite{li2024beyond} and other components of social interactions \cite{dechsling2022virtual}. In contrast, visual guidance should be in an abstract forms rather than realistic objects to assist without excessively occluding the user's view \cite{tan2023mindful}. Our work confirms and contrasts these preliminary recommendations, as discussed in \S 5.3.

\section{Methods}

\subsection{AR Interventions for Emotion Regulation}

Our work creates a repository of \textbf{AR interventions} for real-time emotion regulation. We define AR interventions as AR features---visual, auditory, or environmental---that provide real-time, in-situ support for regulating emotional and psychological reactions while allowing users to remain grounded in their real-world surroundings. Drawing from affective computing and emotion regulation theories \cite{tao2005affective}, we characterize these interventions along four dimensions:
\begin{itemize}
    \item \textbf{Seamless}: AR interventions should not disengage users from their ongoing activities or surroundings, maintaining situational awareness while providing contextually relevant and supportive visual, auditory, or environmental cues \cite{10.1145/3613904.3642123}.

    \item \textbf{Real-Time}: AR interventions should adapt to the user's emotional state \cite{Wang2025} or contextual cues, ensuring that assistance feels timely, relevant, and naturally integrated into ongoing activities while maintaining engagement \cite{Dixon2016}.

  \item \textbf{User-Focused}: AR interventions should be tailored to the users' needs, preferences, and capacities \cite{juarez2017user} and should not feel intrusive or difficult to sustain \cite{conrad2024digitization} in daily life.

  \item \textbf{Oriented Towards Emotion Regulation}: AR interventions should actively scaffold established emotion regulation processes, reducing emotional distress \cite{rolston2017emotion, chen2024emotion}.

\end{itemize}

\begin{figure*}[h]
  \centering
  \includegraphics[width=\linewidth]{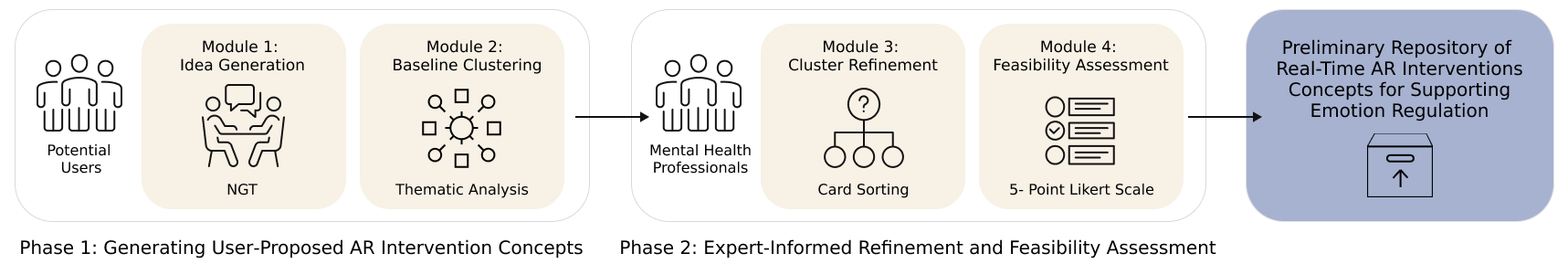}
  \caption{Two-phase participatory process to co-design AR interventions. Phase 1 generates ideas through NGT followed by thematic clustering. Phase 2 refines the clusters through card sorting and a feasibility analysis.}
  \Description{Here the description of the images}
  \label{fig:schematic}
\end{figure*}

 This work follows a participatory design process aligned with our research objectives: (RO1) eliciting a comprehensive set of user-proposed real-time AR interventions concepts, (RO2) organizing these concepts into thematic clusters, and (RO3) assessing the clusters' perceived feasibility  with mental-health professionals.   The process consists of four modules across two phases (Figure~\ref{fig:schematic}).

\subsection{Phase 1: Generating User-Proposed Intervention Concepts}

Phase 1 focused on co-creating real-time AR interventions with anxiety-prone individuals. This phase leveraged two modules to collect both quantitative and qualitative data. The data included the participants' proposed ideas, transcripts of group discussions, and ranking the scores and frequencies of the ideas. The output of this phase was a collection of AR design concepts and an initial set of clusters to validate in the next phase.

\subsubsection{Module 1: Idea Generation}

We employed a variation of the Nominal Group Technique (NGT) to elicit user-proposed ideas for real-time AR interventions (RO1). NGT is a structured method commonly used in healthcare to support decision-making and consensus-building \cite{Cantrill2011, atherton1976group}, as it supports equal participation and minimizes dominance effects \cite{DeRuyter1996}. Following recommendations in the literature \cite{harvey2012nominal}, we conducted NGT sessions with 5--9 participants. Each session followed the stages below to answer a guiding nominal question: \textbf{How can AR Interventions be designed to enhance emotion regulation in real time without disrupting task performance?}

\begin{itemize}
\item \textbf{Contextual Introduction (Stage 1):} Participants watched two videos to establish a shared understanding of the problem space and AR's capabilities. The first video illustrated everyday scenarios of emotional dysregulation. The second video demonstrated a variety of AR capabilities.

\item \textbf{Silent Generation of Ideas (Stage 2):} Participants generated ideas independently in response to the NGT question, using text descriptions or sketches.

\item \textbf{Listing of Ideas (Stage 3):} Participants shared their ideas with the NGT group, one at a time without discussion.

\item \textbf{Discussion of ideas (Stage 4):} Participants collectively discussed the ideas to refine meaning, explore reasoning, and group similar ideas.

\item \textbf{Ranking the Top Ten Ideas (Stage 5):} Each participant selected their ten preferred ideas and assigned scores from 1 to 10, with 10 indicating the highest preference.

\item \textbf{Discussion and Re-Ranking (Stage 6):} Rankings were compiled to define a collective top-ten list. Then, participants discussed the list and re-ranked their ideas independently.

\item \textbf{Re-Ranking based on Cumulative Top-Ten (Stage 7):} Except for the first group, each NGT group discussed the previous group's top-ten list alongside their own. They then re-ranked the ideas, including those from the prior group. This iterative process enabled convergence across groups and produced a cumulative top-ten of prioritized ideas.

\end{itemize}

The NGT scores of each idea were normalized at the group level to quantify their importance. For each session \(s\) and idea \(i\), we computed the summed rank \(\mathrm{RankSum}_{s,i}\) across all participants in \(s\). This value was then divided by the maximum possible score for that session, defined as \(10 \times |P_s|\), where \(|P_s|\) is the number of participants in \(s\) and 10 is the maximum per--participant score. The normalized score for each idea is computed as shown below, where a score of 1 means all participants placed the idea at the top of rank, and a score of 0 means that no participant included the idea in their top:

\[
\mathrm{NormScore}_{s,i} = \frac{\mathrm{RankSum}_{s,i}}{10\times|P_s|}
\]

\subsubsection{Module 2: Baseline Clustering}

We applied Thematic Analysis 
to organize the proposed ideas into clusters \cite{10.1145/3544548.3581203}. We followed the six-phase approach described in \cite{AHMED2025100198}: (1) familiarization with the data, (2) generating initial codes, (3) searching for clusters, (4) reviewing clusters, (5) defining and naming clusters, and (6) reporting.  The analysis was conducted by a single researcher in the team, and the output was evaluated in the next phase. Consistent with Reflexive Thematic Analysis theory \cite{Byrne2022}, codes and clusters were treated as interpretive outcomes of the researcher's engagement with the data rather than objective categories requiring inter-coder agreement.

\subsection{Phase 2: Expert-Informed Refinement and Feasibility Assessment}

Phase 2 used expert feedback to assess the feasibility of the proposed AR interventions and their clusters. Two modules used Phase 1's outcomes to collect quantitative and qualitative data from card-sorting sessions and questionnaires. The output of this phase was an expert-validated set of clusters and their corresponding ideas.

\subsubsection{Module 3: Cluster Refinement}
This module organized the ideas into thematic clusters (RO2). Using Lucidchart \cite{faulkner2018lucidchart}, we conducted an online hybrid card-sorting activity \cite{10933924} that lasted approximately 60 minutes per participant. The activity consisted in categorizing the ideas proposed in Phase 1 into the identified clusters based on which of these described the goal of each idea better. Participants were also allowed to create new categories \cite{10.1007/978-3-030-96960-8_4}, leave cards uncategorized, or assign cards to multiple clusters when appropriate. To assess expert agreement, we applied a consensus rule based on the number of experts agreeing on each categorization, labeling assignments as \textit{Clear}, \textit{Strong}, \textit{Moderate}, \textit{Slight}, or \textit{Weak}. Each card was then classified into the cluster with the highest number of experts votes. In cases where there was a tie, the card was assigned to all tied clusters only if the agreement reached the moderate threshold. Otherwise, cards with tied or low vote counts were classified as \textit{Uncategorized}, reflecting insufficient or ambiguous consensus.

The inter-rater expert agreement was computed through the Krippendorff's $\alpha$ \cite{krippendorff2018content} and Randolph's free-marginal $\kappa$ \cite{randolph2005free}. Since experts could assign a card to multiple clusters, each cluster was treated as a binary membership decision (1 = assigned, 0 = not assigned), and agreement was computed separately for each cluster.

\subsubsection{Module 4: Feasibility Assessment}
The goal of this module was to assess the perceived feasibility of the clusters through expert clinical feedback (RO3). Mental health professionals were recruited to validate the clusters through a 5-point Likert scale questionnaire with five evaluation criteria: \textit{effectiveness} (how well the intervention supports emotion regulation), \textit{functionality} (how practical the intervention is for daily use); \textit{usability} (how understandable and easy the intervention is to use), \textit{clinical grounding} (how well the intervention aligns with evidence-based or clinically supported emotion regulation techniques), and \textit{non-distracting} (how well the intervention supports users without being  disruptive). These dimensions were inspired by prior assessments of AR applications for mental health \cite{bakir2023use} and adapted to our research context.

\section{Results}

\subsection{User-Generated Ideas from NGT Groups}

\subsubsection{Participant Demographics}
Phase 1 recruited 40 participants (29 male, 9 female, 2 non-binary), aged 20--44. A recruitment screening was applied using the Generalized Anxiety Disorder-7 (GAD-7) \cite{Spitzer2006}: a minimum score of 1 was required for participation to encourage ideas grounded in real-life emotion regulation strategies. The GAD-7 scores indicated a broad range of anxiety severity: 16 scored below 5, 17 between 5--10, and 7 between 10--15. 17 participants reported that anxiety significantly impacts their daily life. Regarding familiarity with AR, 25 were ``somewhat familiar'', 9 ``very familiar'', 5 ``not familiar'', and 1 an ``expert''. Confidence in emotion regulation techniques varied: 8 ``very confident'', 20 ``moderately confident'', 11 ``slightly confident'', and 1 ``not at all confident''. The study was approved by our Institutional Review Board (IRB).

\subsubsection{NGT Analysis}
\begin{figure*}[h]
  \centering
  \includegraphics[width=\linewidth]{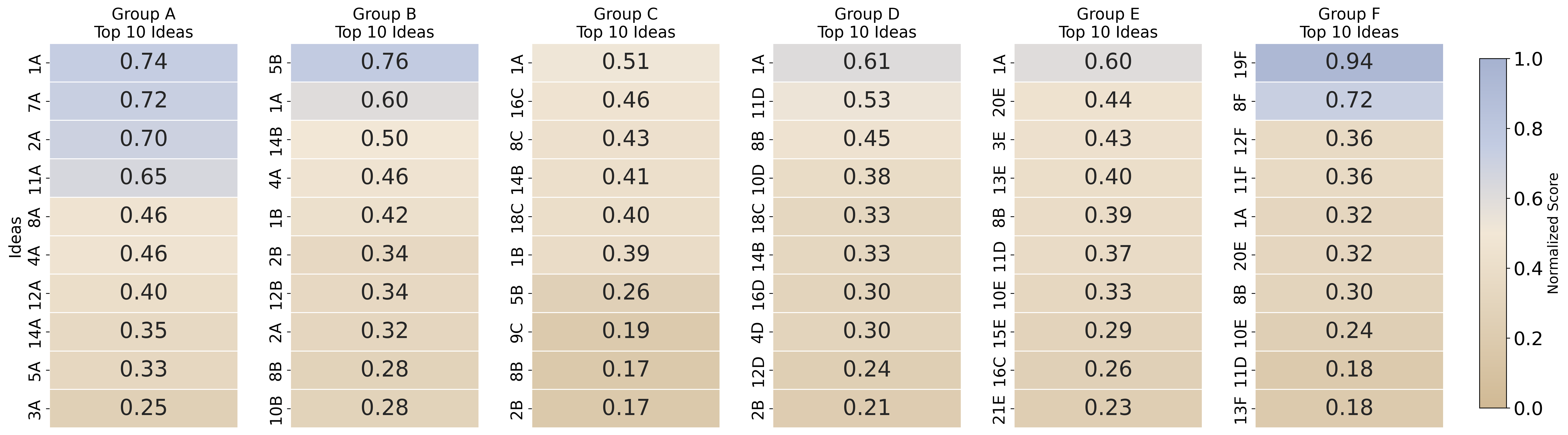}
  \caption{Heatmaps of top ten ideas per group showing normalized consensus scores from NGT sessions. Purple tones indicates stronger group-level agreement across the co-design sessions. Group F showed the strongest consensus, Groups A and B showed high consensus, Groups D and E displayed only moderate consensus, and Group C showed weak consensus.}
  \label{fig:HeatmapTopTen}
\end{figure*}

Six NGT groups (A through F) were run with 8, 5, 7, 8, 7, and 5 participants, respectively. A total of 106 ideas for real-time AR interventions were generated across these groups (A: 15, B: 14, C: 19, D: 12, E: 24, F: 22). Each idea was assigned an ID based on its group. For instance, Idea A10 refers to the 10th generated idea from the NGT session A. Figure~\ref{fig:HeatmapTopTen} shows the normalized scores of the cumulative top-ten for each NGT session. ``\textit{An AR task counter that shows progress/tracks time of tasks and congratulates completion}'' (1A), ``\textit{AR Software that motivates user to increase physical and/or social activity by providing a safe environment of their choosing}'' (5B); ``\textit{Create a virtual pet or replicate the real pet of people as a virtual buddy}'' (7A), and ``\textit{Use AR videos that make you smile and reduce your anxiety}'' (8F) were among the highest ranked ideas. Appendix B shows the full list of proposed AR interventions.

The results show that Group F had the strongest consensus, with Idea 19F receiving a normalized score of 0.94 and Idea 8F scoring 0.72. However, the remaining ideas in Group F showed lower levels of agreement, reflecting greater diversity of opinions. Groups A and B also showed high consensus, with several ideas exceeding 0.70, e.g., 1A, 2A, 7A, and 5B. In contrast, Groups D and E only displayed moderate consensus, with their top ideas reaching scores of about 0.60. Group C showed the lowest overall agreement, with its highest-ranked idea (1A) scoring just 0.51. Few ideas appeared across groups, indicating limited direct convergence. Most notably, Idea 1A was included in the top-ten lists of all six groups and consistently ranked among the highest-scoring ideas. Idea 8B also appeared in five groups, while Ideas 16C and 11D were present in three groups each. The remaining ideas appeared in fewer than three groups, or were unique to a single group.

\subsection{Thematic Analysis}
Our thematic analysis resulted in 42 distinct codes that captured relevant aspects of the proposed ideas. The full list of thematic codes, their description, and the number of times each code appeared among the ideas can be found in Appendix C. Additionally, Appendix D describes how these thematic codes are represented for all the generated AR intervention ideas.

These codes were subsequently organized into nine baseline clusters, as detailed below:

\begin{itemize}
    \item \textbf{Cognitive Scaffolding for Emotional Regulation}: Ideas that guide thinking, memory, or decision-making in ways that support emotional regulation and task completion.
    \item \textbf{Restorative Breaks and Recovery}: Ideas that encourage users to pause, recover, or prevent stress from building up, through short breaks or recovery activities.
    \item \textbf{Managing Sensory and Cognitive Load}: Ideas that focus on reducing distractions or helping users manage information and stimuli, making it easier to maintain focus and mental clarity.
    \item \textbf{Embodied \& Multisensory Regulation}: Ideas that use the body, touch, movement, or multiple senses to help regulate emotions or provide grounding experiences.
    \item \textbf{Emotionally Supportive Environments}: Ideas that involve modifying or adapting the user's environment or surroundings to create a more supportive emotional context.
    \item \textbf{Companionship \& Social Presence}: Ideas that provide a sense of companionship or social connection, either through virtual companions or reminders of supportive relationships.
    \item \textbf{Shaping Positive Appraisals}: Ideas that help users reframe situations, receive encouragement, or build positive interpretations of experiences.
    \item \textbf{Emotion Awareness \& Adaptive Feedback}: Ideas that involve detecting, monitoring, or providing feedback on users' emotions and adapting responses accordingly.
    \item \textbf{Collaborative \& Team-based Support}: Ideas that involve supporting interaction, collaboration, or engagement with others in team or group contexts.
\end{itemize}

Some codes appeared more frequently than others, reflecting recurring user needs. For instance, within the cluster \textit{Emotionally Supportive Environments}, the code \textit{Environmental Modification} was particularly salient, with 23 quotes describing how AR could ``\textit{visually alter the aesthetics of the space to reflect a more familiar area}'' (4A) or ``\textit{create a calming environment around someone who is working in order to relax them and not disrupt tasks}'' (14B). Similarly, codes such as \textit{Social Interaction Support} (cluster: \textit{Collaborative \& Team-based Support}) and \textit{Virtual Companion} (cluster: \textit{Companionship \& Social Presence}) highlighted the need of interpersonal regulation strategies, ranging from structuring teamwork---``\textit{conduct meetings/work-related activities directly in an AR-created environment that helps expressing ideas in detail to avoid unnecessary arguments}'' (10C)---to providing a sense of comfort---``\textit{generating an AR version of a cute animal (e.g., cat, dog, etc.) to play with}'' (4C). Together, these codes and their groupings formed the baseline clusters that structured the subsequent expert card-sorting and feedback.

\subsection{Expert Perceptions of Feasibility for the Clusters and Ideas}

\subsubsection{Participant Demographics}
Phase 2 recruited 10 mental health professionals (8 female, 2 male), aged 18--54. Participants were required to either hold a Master's degree or higher in a relevant field (e.g., psychology, psychiatry, counseling) or have at least two years of professional experience providing mental health care. Of the participants, six held doctoral degrees (PhD, PsyD, or MD), and four held master's degrees. Four participants were fully licensed to practice, four were in training or under supervised practice, and two did not hold a license. Four reported more than 10 years of professional experience, one reported 6--10 years, and five reported 2--5 years. All reported having experience in activities such as individual and group therapy, mental wellness coaching, supervision of clinical assessment, and emotion regulation interventions. None of the participants had prior experience with AR tools in clinical settings. The study was also approved by our IRB.

\subsubsection{Assigning AR Interventions to Clusters}

\begin{figure*}[h]
  \centering
  \includegraphics[width=\linewidth]{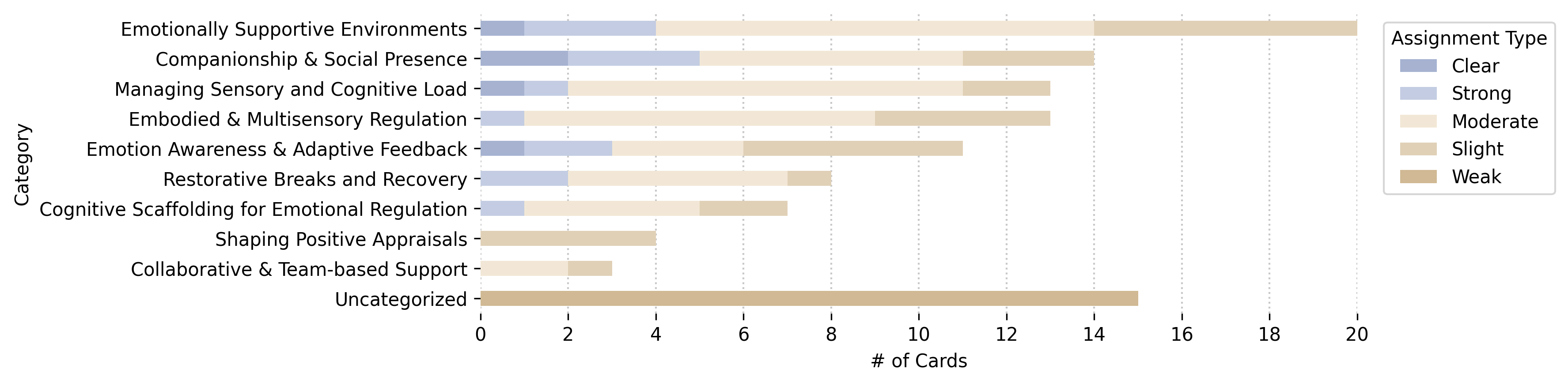}
  \caption{Distribution of card assignments across cluster categories segmented by consensus strength. The horizontal bar chart shows the number of cards assigned to each category, segmented by level: clear, strong, moderate, slight, and weak. Clusters such as \textit{Emotionally Supportive Environments} and \textit{Companionship \& Social Presence} received more cards with higher consensus.}
  \Description{Here the description of the images}
  \label{fig:cards_distribution}
\end{figure*}

Figure~\ref{fig:cards_distribution} shows the final assignments of ideas across agreement levels per cluster, following the criteria in \S 3.3.1. Appendix E showcases how the experts used the card-sorting task to distribute each of the 106 AR intervention ideas among the resulting clusters. Since 10 experts were recruited, the consensus rule was set as \textit{Clear} = 10/10 experts, \textit{Strong} = 8--9/10, \textit{Moderate} = 6--7/10, \textit{Slight} = 4--5/10, or \textit{Weak} $\leq$ 3/10. Several clusters included ideas that reached Clear consensus, such as \textit{Companionship \& Social Presence} (2 ideas), indicating unanimous agreement among experts. Other categories also contained ideas with Strong consensus, such as 3 ideas in both the \textit{Companionship \& Social Presence} and \textit{Emotionally Supportive Environments} clusters. In contrast, 88 out of the 106 idea assignments only reached moderate or lower consensus, suggesting variable expert interpretations.

Inter-rater reliability was computed separately for each category. As shown in Table~\ref{tab:expert-overall}, agreement across categories was limited, with Krippendorff's $\alpha$ ranging from 0.16 to 0.53 and Randolph's $\kappa$ ranging from 0.68 to 0.94. The results suggest which clusters appear more conceptually robust and which may require refinement, consolidation, or clearer definitions. Despite low agreement, no expert suggested the creation of additional clusters.

\subsubsection{Cluster Feasibility Rankings}
\begin{figure*}[h]
  \centering
  \includegraphics[width=\linewidth]{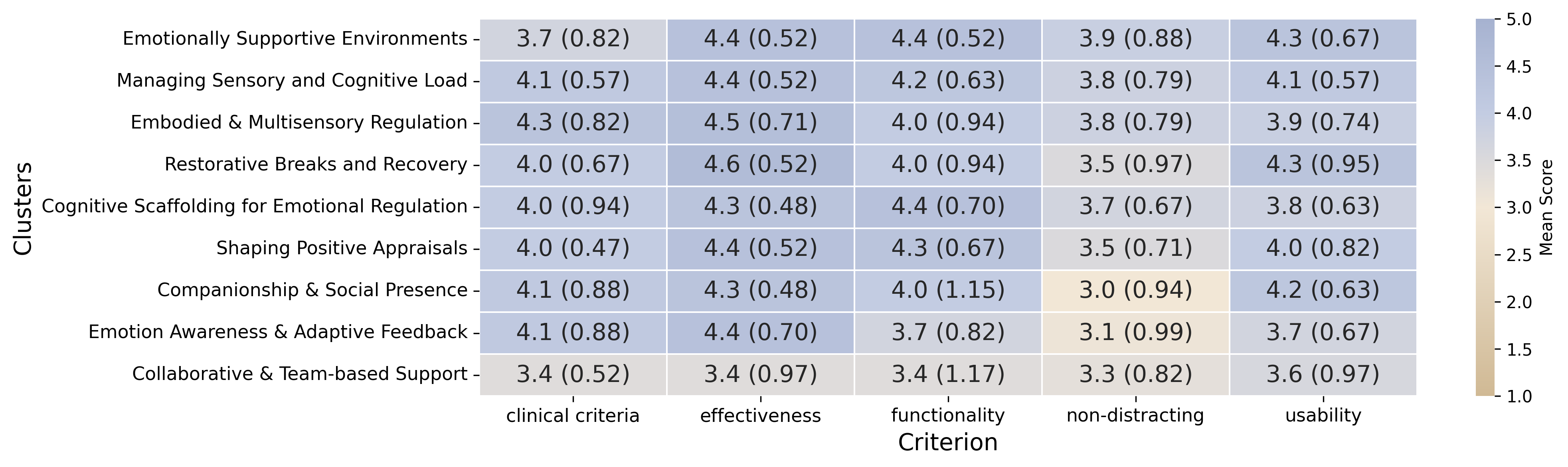}
  \caption{Mean ratings of intervention clusters across five evaluation criteria including clinical and usability measures. Purple tones indicates higher ratings across criteria. Clusters such as \textit{Embodied \& Multisensory Regulation}, \textit{Emotionally Supportive Environments}, and \textit{Managing Sensory and Cognitive Load} received some of the highest ratings across multiple criteria.}
  \Description{Here the description of the images. Descriptive statistics are shown as mean(standard deviation).}
  \label{fig:heatmap}
\end{figure*}

The experts evaluated the baseline clusters feasibility using the 5-point Likert scale questionnaire in Appendix A. The results---summarized across the five criteria as mean and standard deviation (M , SD)---show that professionals rated \textit{Emotionally Supportive Environments} ($M = 4.14$, $SD = 0.32$), \textit{Managing Sensory \& Cognitive Load} ($M = 4.12$, $SD = 0.22$), and \textit{Embodied \& Multisensory Regulation} ($M = 4.10$, $SD = 0.29$) as the most promising approaches, with consistently high scores across criteria. Mid-ranked categories included \textit{Restorative Breaks \& Recovery} ($M = 4.08$, $SD = 0.41$), \textit{Cognitive Scaffolding for Emotional Regulation} ($M = 4.04$, $SD = 0.30$), and \textit{Shaping Positive Appraisals} ($M = 4.04$, $SD = 0.35$). \textit{Collaborative \& Team-based Support} received the lowest overall evaluation ($M = 3.42$, $SD = 0.11$), suggesting limited perceived practicality. Several categories, such as \textit{Companionship \& Social Presence} ($M = 3.92$, $SD = 0.53$) and \textit{Emotion Awareness \& Adaptive Feedback} ($M = 3.80$, $SD = 0.49$), showed greater variability across criteria, indicating mixed professional perspectives. Figure \ref{fig:heatmap} shows a criterion-level analysis to highlight the differences across cluster preferences.

\begin{table}[h]
  \caption{Inter-rater reliability scores by intervention category using Krippendorff's $\alpha$ and Randolph's $\kappa$. Higher values represent high correlation between the rater's responses.}
  \label{tab:expert-overall}
  \begin{tabular}{p{5.5cm} c c}
    \toprule
    \textbf{Category} & \textbf{Kripp $\alpha$} & \textbf{$\kappa$ Free} \\ \midrule
    Cognitive Scaffolding for ER & 0.409 & 0.875 \\
    Restorative Breaks and Recovery & 0.333 & 0.846 \\
    Managing Sensory and Cognitive Load & 0.200 & 0.680 \\
    Embodied \& Multisensory Regulation & 0.293 & 0.759 \\
    Emotionally Supportive Environments & 0.308 & 0.676 \\
    Companionship \& Social Presence & 0.531 & 0.849 \\
    Shaping Positive Appraisals & 0.155 & 0.883 \\
    Emotion Awareness \& Adaptive Feedback & 0.357 & 0.814 \\
    Collaborative \& Team-based Support & 0.353 & 0.936 \\ \bottomrule
  \end{tabular}
\end{table}

\subsection{Preliminary Repository of AR Interventions}

To construct the repository, we applied the consensus rules in \S 3.3.1 to determine which clusters and ideas to include. The threshold used was the moderate level of agreement (6--7/10), resulting in eight clusters from the original nine, as \textit{Shaping Positive Appraisals} did not include ideas above the slight level of agreement (4--5/10). We then systematically reviewed the ideas assigned to each cluster to merge overlapping entries and remove redundancies. Then, we translated the ideas into actionable, design-oriented concepts that can inform AR interventions. The resulting repository, described in Table~\ref{tab:final_repo1}, contains the eight clusters, each containing user-centered, expert-informed design concepts to create AR interventions. These clusters should not be interpreted as rigidly independent; many ideas spanned multiple categories, reflecting the interconnected nature of emotional needs and regulation strategies. Accordingly, the repository is intended as a set of flexible design concepts: clusters can be combined or adapted to address overlapping or multifaceted emotional needs, enabling the development of AR interventions that are both targeted and integrative.

\begin{table*}
  \caption{Preliminary repository of real-time AR interventions for emotion regulation. Each cluster is represented by merged actionable design concepts, derived from the ideas ellicted by the participants in Phase 1.}
  \label{tab:final_repo1}
  \begin{tabular}{p{4cm}p{4cm}p{6cm}}
    \toprule
    \textbf{Cluster} & \textbf{Associated Codes} &\textbf{Merged Actionable Design Concepts (with Derived Cards)} \\
    \midrule

    \textit{Cognitive Scaffolding for Emotional Regulation} &
    Problem-Solving Strategies & AR prompts or overlays that guide users to pause, reflect on a situation, and consider alternative interpretations and solutions.\\
    & Concept Visualization & AR visualizations that translate complex information into interactive visual or short video demonstrations to support understanding.  \\
    \midrule

    \textit{Collaborative \& Team-based Support} &
    Collaborative Workspaces & AR environments that support collaborative work spaces through interactive visual tools for organizing, expressing, and manipulating ideas. \\
    & Shared Gamified Activities & Short AR games that create brief shared breaks during work to reduce stress, restore focus, and strengthen social connection. \\
    \midrule

    \textit{Companionship \& Social Presence} &
    Virtual Companion Agents & AR companions that provide emotional support through responsive and calming interactions based on user behavior or emotional state. \\
    & Guided Grounding Prompts & AR prompts or voice-guided interactions that support emotion regulation through structured grounding exercises (e.g., noticing surrounding objects, sounds, or sensations). \\
    \midrule

    \textit{Embodied \& Multisensory Regulation} &
    Multisensory Feedback & AR experiences that combine multisensory cues (e.g., sound, visuals, vibration, pressure, or temperature) to support emotional regulation through calming or rhythmic feedback.\\
    & Attentional Deployment & AR cues that direct attention to calming sensory elements in the environment (e.g., colors, textures, movement, or natural features) to shift focus and support grounding. \\
    \midrule

    \textit{Emotion Awareness \& Adaptive Feedback} &
    Emotion State Visualization & AR visualizations that display users' emotional states in real time through simple indicators (e.g., color overlays, emotion meters, or interactive panels) to support emotional awareness.\\
    & Adaptive Feedback Prompts & AR systems that monitor physiological or behavioral indicators of stress and provide adaptive calming feedback through interactive cues. \\
    \midrule

    \textit{Emotionally Supportive Environments} &
    Personalized Spatial Aesthetics & AR modifications that transform the surrounding environment into calming or meaningful spaces through personalized spatial overlays. \\
    & Restorative Environments & AR environments that temporarily introduce calming visuals, or ambient cues to help users reset before returning to tasks. \\
    \midrule

    \textit{Managing Sensory and Cognitive Load} &
    Highlighting and Filtering & AR techniques that emphasize task-relevant information and guide attention while suppressing distractors and reducing sensory overload. \\
    & Task Progress & AR visualizations that track task progress and provide real-time feedback. \\
    \midrule

    \textit{Restorative Breaks and Recovery} &
    Guided Micro-Breaks & Short AR experiences that provide brief breaks (e.g., calming visuals, playful interactions, or guided pauses) to support cognitive recovery. \\

    \midrule

  \end{tabular}
\end{table*}

\section{Discussion}

The main objective of this work is to create a comprehensive repository of AR interventions for emotion regulation that can be used as common grounding for the development of future AR applications for mental health. This work cannot be interpreted in a vacuum. Instead, it draws on the extensive literature on AR, mental health, and assistive technologies, and draws connections between the findings in these fields. Throughout this section, parallels are drawn to highlight how existing works can be explained and connected to the principles described in our repository of AR design concepts.

\subsection{Patterns in User-Generated Ideas for AR Interventions}

Our co-design approach was successful in eliciting ideas for real-time AR interventions for emotion regulation, addressing \textbf{RO1}. Prior literature has shown that emotion regulation is inherently personal, and no single technique works equally well for everyone \cite{Kozubal2023}. In line with this, our findings highlight the diversity of perspectives participants brought to the design space of AR interventions. These proposed ideas varied in modality (e.g., interactive dashboards, virtual avatars, environmental overlays), level of disruption to the ongoing task (e.g., seamlessly integrated interventions versus short breaks or minigames that temporarily disengage the user from the primary task), and context of use (e.g., daily activities, meetings, studying, and work settings). These approaches included task management support, virtual pets with varying levels of interactivity, emotion meters to track users' emotional states, guided physical or breathing activities, tracking nervous tics to provide real-time feedback, multisensory feedback, and overlaying natural elements to enhance mood. While some of these ideas have been explored individually in prior work (e.g., \cite{10.1007/978-3-031-05431-0_21, 10.1145/3532525.3532527, 10.1145/3491101.3519874}), others remain less underexplored, such as the use of diminished reality (e.g., hiding overwhelming stimuli or highlighting relevant task objects) for emotional support purposes. Overall, our study shows how participants surfaced these approaches within a unified design space while highlighting important patterns and user needs.

One salient pattern was the desire for personalization and meaningful aesthetics. For instance, when users proposed ideas related to virtual companionship, they expressed them in varied forms: animals ``\textit{a cat that wags its tail with your biometrics or a dog}'' (8B), fictional creatures ``\textit{adding small fictional animals to the surroundings and allowing user interactions}'' (23E), or plants ``\textit{some sort of virtual plant that reacts to your emotions}'' (3C). Similarly, ideas within the \textit{Emotionally Supportive Environments} cluster highlighted the importance of recreating environments tailored to user preferences, including past memories or experiences that evoke positive emotions and a sense of familiarity. Another key design element was interactivity. Ideas such as ``\textit{ AR can be used to give them the feeling of their loved one being in the same room with them}'' (6C) and ``\textit{AR desk pet, does not need active attention but can be interacted with when needed}'' (9B) suggest that, even when minimizing distraction was a requirement, users emphasized the importance of adaptable interaction levels and modalities that can be modulated based on user needs to avoid disrupting ongoing activities. These expressions align with prior research on affective design and computing that emphasizes the importance of emotional connection \cite{schlicher2025emotionally, carpenter2025precision}.

\subsection{Expert Feedback and Clinical Grounding}

Designing AR interventions for mental health requires approaches that balance user perspectives with clinical expertise \cite{CHINSEN2025413, https://doi.org/10.1111/jar.70022}. Without this balance, there is a risk of unintentionally supporting maladaptive behaviors that may not foster long-term resilience or effective emotion regulation skills \cite{dixon2016treatment, Chaaya2025}. While user expectations play an essential role in fostering emotional connection and engagement, the involvement of mental health professionals is critical for assessing the feasibility and suitability of user-generated ideas.
To address \textbf{RO2} and \textbf{RO3}, we organized the generated ideas into thematic clusters and evaluate their feasibility through expert feedback. Following this, our co-design results indicate that although experts were open to a diverse range of ideas, they consistently favored the clusters \textit{Emotionally Supportive Environments}, \textit{Managing Sensory \& Cognitive Load}, and \textit{Embodied \& Multisensory Regulation} as the most promising across evaluation criteria.
In contrast, participants frequently proposed ideas within \textit{Companionship \& Social Presence} and \textit{Emotionally Supportive Environments}---such as virtual pets and natural overlays---seeking more human-like interactions and soothing environments. This tendency aligns with prior psychology research noting that, in the absence of social interactions, individuals seek technology-mediated alternatives to fulfill their emotional needs \cite{HERBENER2025103409}. However, experts raised concerns about the potential distraction of these strategies, as animated companions may capture attention and shift focus away from the primary task, encouraging task avoidance behaviors. This perception arose from poor design practices that treat animations as decorative elements that hinder performance \cite{munoz2023functionalanimation}. Nevertheless, prior work by Almaral et al. shows that animations can enhance user experience without causing distraction when adapted to the user's context \cite{10.1145/3591156.3591180}. Likewise, the cluster \textit{Restorative Breaks \& Recovery} raised similar concerns, with one professional cautioning that while brief activities (e.g., short games) may help restore focus, without re-grounding mechanisms they risk promoting disengagement from the primary task. This aligns with research showing that video games can serve as coping strategies by providing temporary relief from stress, but may also function as escapist behaviors that lead to disengagement and negative outcomes \cite{sinha2024problematicgaming}. Taken together, these findings highlight the importance of balancing emotional engagement---through interactivity and animation---with non-disruptive design to ensure the practical applicability of these clusters in everyday contexts.

\subsection{Comparisons against existing AR for mental health recommendations}

Building on the defined clusters, we connect our findings to design guidelines from HCI, telehealth, and chatbots discussed in Section 2.3. These guidelines emphasize agency, non-intrusiveness, and contextual integration. Within our repository, the cluster \textit{Managing Sensory \& Cognitive Load} embodies contextual integration by adapting interventions to the user's environment, emphasizing task-relevant information while suppressing distractions. Likewise, the cluster \textit{Cognitive Scaffolding for Emotion Regulation} offers contextualized, real-time AR cues that help users stay organized and prioritize activities without disrupting ongoing tasks. These strategies align with problem-focused coping, which manages stressors through structured, goal-directed behaviors \cite{schoenmakers2015copingloneliness}. Although such approaches have been explored in performance-focused studies \cite{10.1145/3706598.3713415, 8862073}, their application to mental health outcomes remains underexplored.

Beyond these principles, prior work highlights the importance of customization, social interaction, and multisensory immersion \cite{marto2024scope, dechsling2022virtual}. In line with this, the \textit{Embodied \& Multisensory Regulation} cluster combines tactile, auditory, and visual cues to foster grounding interactions, which have been shown to reduce cognitive load and support emotional regulation \cite{Alwashmi2023Enhancing,Marucci2021The}. Similarly, the clusters \textit{Emotionally Supportive Environments} and \textit{Companionship \& Social Presence} reflect customization by incorporating personal or familiar cues within the AR space, creating experiences tailored to users' preferences. However, in contrast to prior work that emphasizes enhancing the real world with calming overlays \cite{viczko2021effects} and companion-based supports \cite{norouzi2019walking}, our findings suggest that these approaches may introduce distraction when integrated into ongoing daily activities. Therefore, careful modulation of these interventions is necessary to ensure seamless integration without disrupting the user's primary tasks \cite{firth2017challenges}. Moreover, \textit{Companionship \& Social Presence} and \textit{Collaborative \& Team-based Support} clusters emphasize social interaction through virtual companions and shared gamified activities, fostering presence and empathy while reducing loneliness and enhancing motivation. Prior research suggests that humans tend to interact with computers similarly to how they interact with other humans \cite{Merrill15032022, reeves1996media}. Thus, social presence and warmth features in computer agents can provide outcomes comparable to supportive social networks \cite{Merrill15032022}. For instance, AI chatbots can foster a sense of being heard and understood through more empathic interactions, reducing loneliness and improving well-being \cite{kim2025socialchatbots}.

While the clusters are derived from AR-based interventions, their relevance extends to other immersive technologies, including MR and VR, warranting consideration of their transferability. Overall, a large portion of the ideas in the repository can be transferred to MR, as long as the interventions remain non-disruptive, seamless, and real-time. For instance, visualizations such as task counters and progress bars, or overlays of familiar figures to reduce loneliness, are not limited to a specific device or technique. Nonetheless, some recommendations can be more effectively delivered by leveraging the affordances of specific technologies. For example, AR is particularly well-suited for use cases that primarily add content to the environment rather than modify it, such as overlays that introduce calming natural elements. In contrast, current MR techniques that rely on pass-through modes often struggle to accurately render nearby screens \cite{bailenson2024seeing}. As such, interventions that require to situate content near screens will work better in AR to prevent visualization artifacts and potential motion sickness. Additionally, applications that aim to reduce overload by suppressing distractors and clutter can benefit from MR's affordances. These applications require techniques such as Diminished Reality \cite{cheng2022towards}, which are easier to implement with MR devices that allow introducing rendering delays. Finally, it is important to keep in mind that VR/AR/MR is a continuum, not discrete, disconnected categories \cite{azuma2002recent}. Depending on the actual AR intervention that needs to be created and the hardware to use, the recommendations from the repository might apply better as fully AR, fully VR, or as a combination.

\subsection{Future Research Directions}

Our thesis is that immersive technologies will continue to expand within mental health contexts and become increasingly ubiquitous, making it essential to guide their design thoughtfully to ensure they support everyday well-being. Accordingly, future research should focus not only on developing immersive technologies that are usable, acceptable, and adaptable to diverse real-world contexts \cite{yang2023immersive}, but also on establishing sustainable approaches for their seamless integration into practice \cite{virtualworlds3030020}. Rather than assuming these technologies should be universally incorporated into mental health care, it is critical to understand when, for whom, and under what conditions immersive interventions are appropriate, as well as how to deploy them ethically and equitably. Building on our findings and preliminary repository of real-time AR interventions, we therefore identify concrete research directions for prototyping and empirical validation.

\begin{itemize}

\item \textbf{Promising clusters for prototyping}: \textit{Embodied \& Multisensory Regulation} (e.g., multisensory, haptic and thermal feedback), \textit{Emotionally Supportive Environments} (e.g., calming natural overlays), and \textit{Managing Sensory and Cognitive Load} (e.g., context-aware filtering and sensory modulation) showed high scores across criteria, making them strong candidates for early deployment in real-world contexts.

\item \textbf{Noticeable gaps in current AR interventions}: While prior AR interventions have explored clusters such as \textit{Emotionally Supportive Environments} and \textit{Companionship \& Social Presence}, they have often been approached without considering their seamless integration with ongoing tasks. As a result, these interventions risk being disruptive. This reveals challenges related to balancing interactivity with integration, user agency, control, and effective modulation in real-world contexts. Addressing these gaps is critical for advancing AR interventions beyond decorative overlays.

\item \textbf{Increasing ecological validity}: Future evaluations of AR interventions should be performed in real-world contexts such as study sessions, workplace, and therapy-adjacent routines. These evaluations should emphasize contextual sensitivity (e.g., distraction levels, social acceptability, ecological validity, and activity type), and link outcomes to emotion regulation and task performance. This line of work will establish evidence-based claims to confirm that AR interventions can provide real-time support in everyday settings without disrupting users.

\end{itemize}

\subsection{Study Limitations}
This work represents a first step towards the development of seamless real-time AR interventions for mental health. Due to its preliminary nature, the repository requires further validation to determine their effectiveness and feasibility. For instance, the impact of the AR interventions may vary across contexts, populations, and use cases, including differences in cultural background, clinical needs, and technological familiarity. Additionally, constraints in real-world settings may affect adoption, e.g., limited time, attention, and device availability. Future work should include longitudinal and ecological validation to assess how these interventions perform in everyday environments.

Moreover, limitations related to participant sampling and expertise were present in both phases. In Phase 1, although our NGT sessions included a diverse cohort of individuals with varying levels of anxiety, the sample was primarily drawn from a university setting and was relatively limited in terms of age range, cultural background, and mental health history. These factors can shape distinct patterns of emotion regulation preferences, and thus our findings represent only a subset of the broader diversity of human emotional experiences and needs. Likewise, in Phase 2, although the number of expert participants (n=10) aligns with prior co-design work in mental health technology \cite{OspinaPinillos2019MentalHealthEClinic,Vial2023MentallysCoDesign}, differences in expertise and professional background may have influenced evaluations of the proposed clusters. For instance, four participants were in training or under supervised practice, and their assessments may differ from those of more experienced clinicians, reflecting differences in experience. Moreover, differences between practitioner- and research-oriented professionals may impact the perceived effectiveness of the ideas. Future work should therefore explore how preferences generalize across populations and include prototype development and validation in clinical contexts.

\section{Conclusion}
In this paper, we present the findings and co-design process used to develop a user-centered and expert-informed preliminary repository of AR design concepts for real-time emotion regulation. Our work addresses the lack of user-centered design resources for real-time AR interventions, which has hindered the development of reusable and clinically grounded solutions in this space. To bridge this gap, we conducted a two-phase participatory design process. In the first phase, we conducted six NGT groups with anxiety-prone individuals (N=40), where participants generated ideas on how AR capabilities could support real-time emotion regulation. In the second phase,  mental health professionals (N=10) organized these ideas into thematic clusters and assessed their feasibility against clinical criteria. This process resulted in eight clusters with actionable design concepts that can guide designers and researchers in developing AR interventions for emotion regulation. Our research contributes to HCI by providing a preliminary repository of reusable design concepts grounded in user perspectives and clinical expertise. This contribution represents a step toward more effective design guidelines for AR-based mental health support. 

\bibliographystyle{ACM-Reference-Format}
\bibliography{sample}

\appendix

\end{document}